\documentclass[aps,prd,twocolumn,superscriptaddress,showpacs,showkeys]{revtex4-1}  
\usepackage{graphicx,color} 
\usepackage{xcolor}
\usepackage{rotating}
\usepackage{amssymb}
\usepackage{amsfonts}
\usepackage{amsmath}
\usepackage{xspace}
\usepackage{mathtools} 

\usepackage{listings}
\makeindex

\usepackage[thinlines]{easytable}

\usepackage[super]{nth}
\usepackage{braket}

\newcommand{\be}{\begin{equation}}
\newcommand{\ee}{\end{equation}}
\newcommand{\bea}{\begin{eqnarray}}
\newcommand{\eea}{\end{eqnarray}}

\newcommand\ie{\mbox{\textit{i.\,e.}}\xspace}
\newcommand\cf{\mbox{c.\,f.}\xspace}
\newcommand\eg{\mbox{e.\,g.}\xspace}

\newcommand\D{\mathrm{d}}

\newcommand\Ord{\mathcal{O}}

\newcommand\op{\hat{\mathcal{O}}}
\newcommand\hil{\mathcal{H}}

\begin{document}

\title{Asymptotic Generalized Extended Uncertainty Principle}





\author{Mariusz P. D\c{a}browski}
\email{Mariusz.Dabrowski@usz.edu.pl}
\affiliation{Institute of Physics, University of Szczecin, Wielkopolska 15, 70-451 Szczecin, Poland}
\affiliation{National Centre for Nuclear Research, Andrzeja So{\l}tana 7, 05-400 Otwock, Poland}
\affiliation{Copernicus Center for Interdisciplinary Studies, Szczepa\'nska 1/5, 31-011 Krak\'ow, Poland}

\author{Fabian Wagner}
\email[]{fabian.wagner@usz.edu.pl}
\affiliation{Institute of Physics, University of Szczecin, Wielkopolska 15, 70-451 Szczecin, Poland}
\date{\today}
\begin{abstract}
We present a formalism which allows for the perturbative derivation of the Extended Uncertainty Principle (EUP) for arbitrary spatial curvature models and observers. Entering the realm of small position uncertainties, we derive a general asymptotic EUP. The leading \nth{2} order curvature induced correction is proportional to the Ricci scalar, while the \nth{4} order correction features the \nth{0} order Cartan invariant $\Psi_2$ (a scalar quadratic in curvature tensors) and the curved space Laplacian of the Ricci scalar all of which are evaluated at the expectation value of the position operator \ie the expected position when performing a measurement. This result is first verified for previously derived homogeneous space models and then applied to other non-trivial curvature related effects such as inhomogeneities, rotation and an anisotropic stress fluid leading to black hole "hair". 

Our main achievement combines the method we introduce with the Generalized Uncertainty Principle (GUP) by virtue of deformed commutators to formulate a generic form of what we call the Asymptotic Generalized Extended Uncertainty Principle (AGEUP).
\end{abstract}


\maketitle

\section{Introduction}
\label{intro}

The standard uncertainty principle of quantum mechanics in its fundamental form does not take into account effects which are expected to arise from an underlying theory of quantum gravity. Taking inspiration from string theory \cite{Amati87,Kempf95}, the Heisenberg uncertainty principle was generalized by the inclusion of the gravitational photon-electron interaction (acceleration) leading to the Generalized Uncertainty Principle (GUP) \cite{Maggiore93,Adler01,GUPReview,Sparsity}. 

As usually assumed, the GUP takes into account the gravitational uncertainty of position related to the minimum fundamental length scale in physics. However, it may not be the only gravitationally induced change. In fact, the curvature of space-time does exert an influence over quantum mechanical uncertainty relations. This is the regime of the Extended Uncertainty Principle (EUP) \cite{Mignemi2010,Anrade,Mureika,EUPThermod} which takes into account the uncertainty related to the background space-time. 

Both components can be formulated in terms of the standard deviations of position $x$ and  momentum $p$
\be
\sigma_x^2=\braket{\hat{x}^2}-\braket{\hat{x}}^2 \hspace{0.5cm} 
\sigma_p^2=\braket{\hat{p}^2}-\braket{\hat{p}}^2 \nonumber
\ee
and combined to yield the most general Generalised Extended Uncertainty Principle (GEUP)  \cite{AdlerDuality,Bambi2008}  
\be
\sigma_x \sigma_p \geq \frac{\hbar}{2} \left(1 + \frac{\alpha_0 l_p^2}{\hbar^2} \sigma_p^2 +  \frac{\beta_0}{r_c^2} \sigma_x^2\right),
\label{GEUP} 
\ee
where $l_{p}$ denotes the Planck length, $\hbar$ the Planck constant, $r_c$ some curvature scale related to the background space-time and $\alpha_0$ and $\beta_0$ the GUP- and EUP-parameters, respectively. 

The GEUP as given by (\ref{GEUP}) has its heuristic Newtonian analogue \cite{Bambi2008} which represents the acceleration of an electron induced by both the gravitational interaction of a photon of energy $E$ and the cosmological Hubble horizon $r_{H} = c/H = (\Lambda/3)^{1/2}$ of de Sitter space 
\be
\ddot{\vec{r}} = \left( - \frac{G(E/c^2)}{r^2} + \frac{\Lambda c^2}{3} r \right) \frac{\vec{r}}{r} ,
\ee
with the cosmological constant $\Lambda,$ the photon-electron distance $r,$ the Hubble parameter $H,$ the gravitational constant $G$ and the speed of light $c$ (the latter two will be set equal to $1$ throughout the remainder of this paper). 

A derivation of the EUP based on the notion of geodesic balls was performed in Refs.  \cite{Schuermann2009,Schuermann2018}. It reflects the influence of spatial curvature on quantum-mechanics on 3-dimensional spacelike hypersurfaces of space-time. The method was applied to homogeneous and isotropic geometries of constant curvature $K$ and the corresponding EUP was calculated.  

In our recent paper \cite{RF2019} we applied this method to calculate the EUP for Rindler and Friedmann horizons and  obtained  corrections to the Hawking temperature and Bekenstein entropy of black holes. In said derivation emphasis was put on the local observer in an accelerated frame or at the center of symmetry seeing the effects related to such a specific choice of frame. Yet, a coordinate-independent (covariant) relation for the EUP was the same as derived earlier in \cite{Schuermann2018}. This was further commented on in Ref. \cite{Sch2020}.  

The purpose of this paper is to derive a general expression for the EUP in the case of small position uncertainties (to be specified below) and thus consider non-trivial (e.g. non-homogeneous) effects of curvature on the EUP. 

The paper is organized as follows. Section \ref{EUPgeom} outlines a method of deriving the EUP as it was applied to homogeneous (constant curvature) spaces in Ref. \cite{Schuermann2018}. In Section \ref{WEUP} we present our perturbative approach to apply it further to small position uncertainties in Section \ref{gensol}. Moreover, in Section \ref{applic} several types of perturbations are discussed to provide exemplary applications of the asymptotic EUP to various geometries of non-trivial curvature. Finally, Section \ref{Summary} is intended to summarize our results.  

\section{Background geometry determined EUP}
\label{EUPgeom}

The reasoning behind Refs. \cite{Schuermann2009,Schuermann2018} restricts our perspective to spacelike hypersurfaces of the underlying space-time (it is impossible to even define a mathematically sound uncertainty relation in standard quantum mechanics otherwise). Thus, as other effects which combine quantum mechanics and gravity, this method is observer dependent. 

Consider a free wave function $\psi$ living on said Riemannian manifold but confined to a geodesic ball ($B_{\rho}$) of radius $\rho.$ Due to their invariance under diffeomorphisms, geodesic balls are the natural generalization of the Heisenberg slit to curved manifolds where $\rho$ corresponds to the slit width thus providing a measure of position uncertainty. Confinement to this domain is ensured by imposing Dirichlet boundary conditions on the wave function.

As the Laplace-Beltrami-operator $\Delta$ (representing the squared momentum operator $\hat{p}^2=-\hbar^{2}\Delta$) is hermitian with respect to the measure $\D\mu$ of the Hilbert space in question $\hil=L^2(B_{\rho}\subseteq{\rm I\!R^3,}\D\mu)$ and thus its eigenvalues provide an orthonormal base of the latter, deriving the uncertainty relation basically boils down to solving the eigenvalue problem
\begin{align}
\Delta \psi + \lambda \psi	&=0~\text{within~}B_{\rho}\label{evp}\\
\psi										&=0~\text{on~}\partial B_{\rho}.
\end{align}

Choosing the wave function to be real (the eigenvalue problem is the same for the real and the imaginary part), the Dirichlet boundary conditions ensure that $\braket{\hat{p}}=0.$ Hence, the momentum uncertainty reads
\begin{align}
\label{Dpgen}
\sigma_p=\sqrt{\braket{\hat{p}^2}} =\hbar\sqrt{-\braket{\psi|\Delta|\psi}} \geq \hbar\sqrt{\lambda_1}
\end{align}
with the \nth{1} eigenvalue of the eigenvalue problem \eqref{evp}.

Multiplying by $\rho,$ the uncertainty relation is obtained. In Ref. \cite{Schuermann2018} it was found that the uncertainty relation for Riemannian 3-manifolds of constant curvature $K$ reads
\begin{align}
\label{Kgeom}
\sigma_p \rho\geq \pi\hbar \sqrt{1-\frac{K}{\pi^2}\rho^2} .
\end{align}
Note that the uncertainty relation derived this way is not of the same kind as the one described by \eqref{GEUP} because it features the characteristic length of confinement $\rho,$ a generalisation of Heisenberg's slit width. Thus, $\rho$ should rather be interpreted as uncertainty and does not represent the standard deviation of position.

\section{Weak curvature EUP formalism} 
\label{WEUP}

Consider a generic weak curvature effect on spacelike 3D-hypersurfaces, \ie for a spatial metric splitting
\begin{align}
\D s^2=\left[g^{(0)}_{ij}+\epsilon g^{(1)}_{ij}+\epsilon^2 g^{(2)}_{ij}\right]\D x^i \D x^j ,
\label{splitting}
\end{align}
where $g^{(0)}_{ij}$ denotes the flat space Riemannian metric in some set of coordinates $x^i$, $i=1, 2, 3$.  The perturbation $\epsilon$ will be treated at \nth{2} order throughout this section. Correspondingly, the inverse metric can be approximated as
\begin{widetext}
\begin{align}
\left[g^{(0)}_{ij}+\epsilon g^{(1)}_{ij}+\epsilon^2 g^{(2)}_{ij}\right]^{-1}		
&=g^{ij}_{(0)}-\epsilon  g^{ik}_{(0)}g^{jl}_{(0)}g^{(1)}_{kl}+\epsilon^2\left[g^{ik}_{(0)}g^{ml}_{(0)}g^{(1)}_{kl}g^{(1)}_{mn}g^{nj}_{(0)}-g^{ik}_{(0)}g^{jl}_{(0)}g^{(2)}_{kl} \right]+\mathcal{O}\left(\epsilon^3\right)\\ \nonumber
																			&=g^{ij}_{(0)}-\epsilon g_{(1)}^{ij}+\epsilon^2\left[g_{(1)}^{ik} g_k^{(1)j}- g_{(2)}^{ij}\right]+\mathcal{O}\left(\epsilon^3\right) ,
\end{align}
\end{widetext}
where the last line was just given for notational reasons, \ie to show that the unperturbed metric may be used to rise and lower indices. Note that the sub- and superscripts $(0),$ $(1)$ and $(2)$ denote the perturbation order and should not be understood as covariant or contravariant indices. 

In order to obtain an EUP for the metric (\ref{splitting}), a general construction of geodesic balls in the weak curvature regime has to be given. Furthermore, the corresponding eigenvalue problem has to be solved in these domains. This will be discussed in the following subsections.


\subsection{Geodesic balls\label{geoball}} 

In order to solve the given quantum mechanics problem, the domain to which the corresponding wave function $\psi$ has to be confined needs to be found. As it is 3-diffeomorphism invariant, the most natural generalisation of Heisenberg's slit width is a geodesic ball. This object is defined by its boundary on which every point $p\in \partial B_{\rho}$ has a constant geodesic distance $\sigma(p_0,p)$ from its center $p_0.$ Physically, the point $p_0$ describes the expectation value of the position operator. 

Fixing $p_0,$ the biscalar $\sigma(p_0,p)$ becomes a scalar field $\sigma(p)$ (related to Synge's world function \cite{Synge1960}) which measures the distance along the shortest geodesic connecting the points $p_0$ and $p.$ Hence, it has to satisfy the differential equation \cite{Kothawala:2014tya}
\begin{align}
g^{ij}\partial_i\sigma\partial_j\sigma=1
 \label{def_geo_dist} 
\end{align}
with the boundary condition $\sigma(p_0)=0.$ 

\begin{figure}[!htb]
\centering
\includegraphics[width=\linewidth]{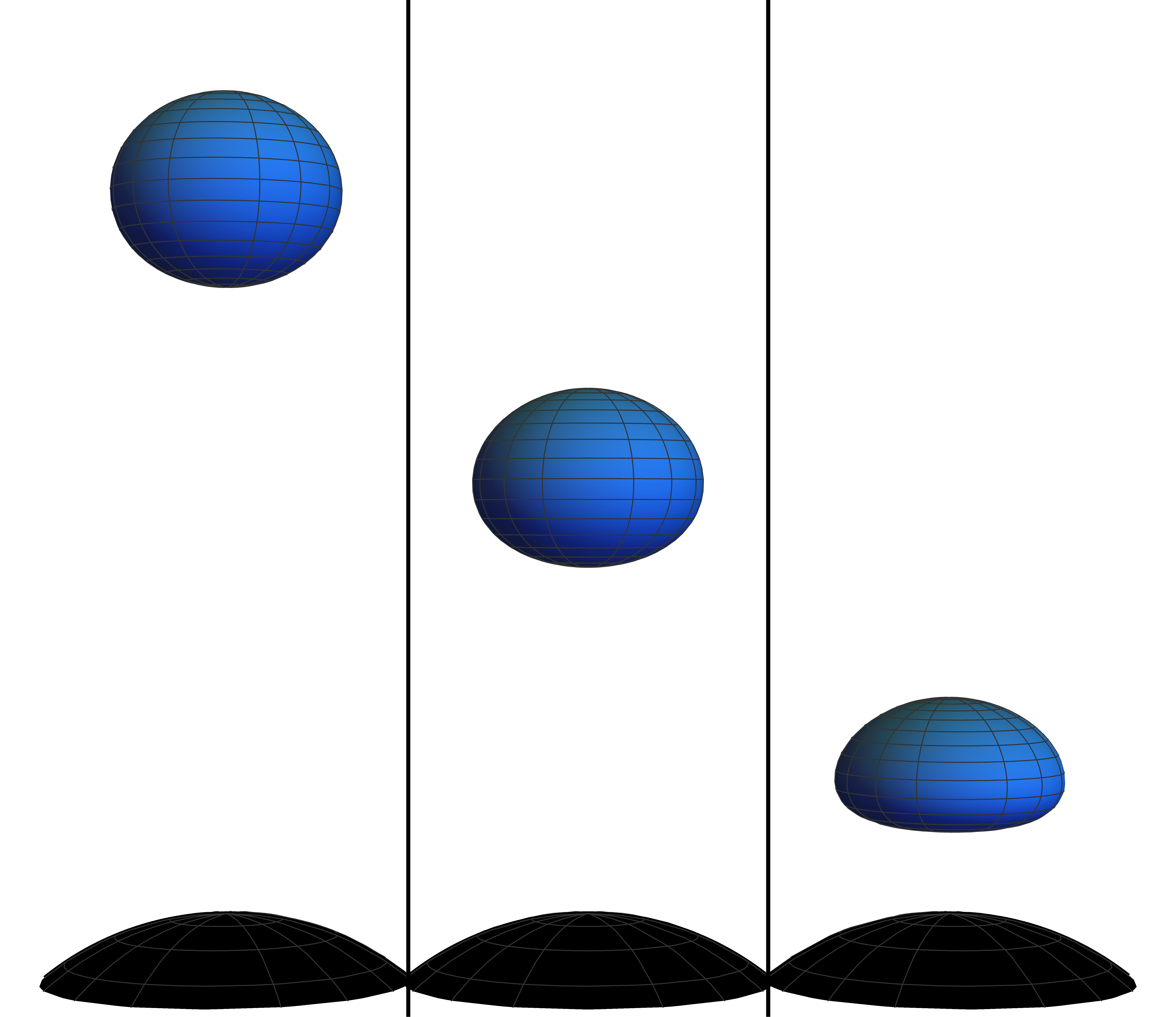}
\caption{Three geodesic balls whith equal geodesic radius $\rho=.4R_S$ but different distances from the center of symmetry $r_0=3.5R_S$ (left), $r_0=2.5R_S$ (mid) and $r_0=1.5R_S$ (right) in a background spatial section of the Schwarzschild static patch characterized by the Schwarzschild radius $R_S$ and described in Schwarzschild coordinates. Surfaces of geodesic balls are coloured blue while black hole horizons are depicted (only for illustration purposes) as black sections of spheres. \label{fig:balls}}
\end{figure}

As it conveniently describes surface normals of the spheres limiting the geodesic balls, we will interpret this scalar as coordinate. These geodesically spherical boundaries can then be conveniently defined by the relation $\sigma=\rho.$ Note that the shape of geodesic balls depends on the coordinates and the background geometry which are chosen and does not need to even closely resemble ordinary balls in flat space and Cartesian coordinates. An example of their change is given in figure \ref{fig:balls}. In this illustration the background consists of constant time slices of the Schwarzschild static patch described by Schwarzschild coordinates. In order to illuminate the distortion, three geodesic balls of equal geodesic radius $\rho$ but different coordinate distances $r_0$ between their center ($p_0$) and the center of symmetry of the space (origin) are compared. The corresponding calculations were done numerically thereby not invoking the small ball approximation made below. It can be seen quite clearly that the distortion of the geodesic balls increases with increasing curvature \ie decreasing coordinate distance from the center of symmetry.

The coordinates which are normal to $\sigma$ can be understood as surface parameters of geodesic spheres. In this paper we denote these as $\chi$ and $\gamma$ to construct the geodesic coordinate system $\sigma^i=(\sigma,\chi,\gamma).$ Imposing orthogonality as motivated above, the "angular" coordinates have to satisfy
\begin{align}
g^{ij}\partial_i\sigma\partial_j\chi=g^{ij}\partial_i\sigma\partial_j\gamma=0\label{ortho}.
\end{align}
From a geometric point of view these coordinates by definition describe the background in such a way that the geodesic balls are not distorted. This might be quite unnatural for arbitrary metrics but simplifies a perturbative treatment in quantum mechanics enormously.

Linearising \eqref{def_geo_dist} and \eqref{ortho} as indicated above, we can write
\begin{align}
\sigma^i\simeq\sigma_0^i+\epsilon\sigma_1^i+\epsilon^2\sigma_2^i\label{geod_coord}
\end{align}
with the geodesic coordinates to $\text{n}^{\text{th}}$ order $\sigma^i_n.$ Similarly, we perturbatively construct derivatives with respect to the employed coordinates as
\begin{align}
\frac{\partial}{\partial\sigma^i}\simeq	&\frac{\partial}{\partial\sigma_0^i}-\epsilon \frac{\partial \sigma_1^j}{\partial \sigma_0^i}\frac{\partial}{\partial \sigma_0^j}\nonumber\\
																&+\epsilon^2\left(\frac{\partial \sigma_1^j}{\partial \sigma_0^i}\frac{\partial \sigma_1^k}{\partial \sigma_0^j}\frac{\partial}{\partial \sigma_0^k}-\frac{\partial \sigma_2^j}{\partial \sigma_0^i}\frac{\partial }{\partial \sigma_0^j}\right)
\end{align}
which works up to \nth{2} order because
\begin{align}
\frac{\partial\sigma^j}{\partial\sigma^i}	&=\delta^j_i+\mathcal{O}(\epsilon^3).
\end{align}

As a starting point for the perturbative quantum mechanical treatment we have to express the given metric in terms of geodesic coordinates. Then \eqref{def_geo_dist} and \eqref{ortho} become simply
\begin{align}
g_{\sigma i}=\delta_i^\sigma\label{geod_met_char}
\end{align}
while the remaining four components of the metric contain all the information about the underlying space.

If the background is flat (as will be assumed later), this can be interpreted as a quasi-spherical (\ie spherical up to corrections of order $\epsilon$) coordinate system.  Then the unperturbed coordinates $\sigma_0^i=(\sigma_0,\chi_0,\gamma_0)$ in \eqref{geod_coord} furnish a spherical coordinate system constructed around $p_0.$ 

For the remainder of this section we assume the metric \eqref{splitting} to be given in terms of geodesic coordinates.

\subsection{Weak field approximation and gravitational waves}

Our geodesic coordinates \eqref{geod_coord} are reminiscent of the weak field approximation applied in the case of gravitational waves \cite{schutz}. The coordinate systems chosen in both cases are similar. In fact, the coordinate describing the geodesic direction of propagation of the gravitational waves is analogous to the geodesic distance coordinate employed in this paper. The other two coordinates are normal to the geodesic motion \ie they are extended along hypersurfaces of constant geodesic distance and can thus be considered analogous to our "angular coordinates". 

However, the representation which is usually used to describe gravitational waves is given in the so-called transverse-traceless (TT) gauge which provides four conditions in the 4-dimensional Lorentzian system while in our 3-dimensional approach there are only three conditions. Therefore the metric is not traceless in the geodesic coordinates defined above - it is merely expressed in a transverse gauge.

\subsection{Eigenvalue problem\label{EP}} 

As in the original approach by Sch\"urmann \cite{Schuermann2018}, the eigenvalue problem to be solved reads
\begin{align}
\left(\Delta+\lambda_n\right)\psi_n	&=0 & \text{inside}~ B_{\rho} ,\\
\psi_n												&=0	&	\text{on}~\partial B_\rho ,	\label{boundarycond}
\end{align}
where $\lambda_n$ and $\psi_n$ ($n$ stands for $nlm$ in 3D) describe the eigenvalues and -functions (there are countably infinitely many) and $\Delta$ denotes the Laplace-Beltrami operator which has to be evaluated on the curved background on which it reads
\begin{align}
\Delta=g^{ij}\left(\partial_i\partial_j-\Gamma^{l}_{ij}\partial_l\right).
\end{align}
Subjecting the background manifold to the metric splitting \eqref{splitting}, the Christoffel symbols become (up to \nth{2} order)
\begin{align}
\Gamma_{ij}^k\simeq\prescript{(0)}{}{\Gamma}_{ij}^k+\epsilon \prescript{(1)}{}{\Gamma}_{ij}^k+\epsilon^2 \prescript{(2)}{}{\Gamma}_{ij}^k
\end{align}
with the unperturbed symbol $\prescript{(0)}{}{ \Gamma}^{k}_{ij}$ and the higher order expressions
\begin{widetext}
\begin{align}
\prescript{(1)}{}{\Gamma_{ij}^k}=\frac{1}{2}	&\left[g_{(0)}^{kl}\left(\partial_i g^{(1)}_{jl}+\partial_j g^{(1)}_{il}-\partial_l g^{(1)}_{ij}\right)-g_{(1)}^{kl}\left(\partial_i g^{(0)}_{jl}+\partial_j g^{(0)}_{il}-\partial_l g^{(0)}_{ij}\right)\right] ,\\
\prescript{(2)}{}{\Gamma_{ij}^k}=\frac{1}{2}	&\Big[	g_{(0)}^{kl}\left(\partial_i g^{(2)}_{jl}+\partial_j g^{(2)}_{il}-\partial_l g^{(2)}_{ij}\right)+\left(g_{(1)}^{km}g^{(1)l}_{m}-g_{(2)}^{kl}\right)\left(\partial_i g^{(0)}_{jl}+\partial_j g^{(0)}_{il}-\partial_l g^{(0)}_{ij}\right)\nonumber\\
																		&-g_{(1)}^{kl}\left(\partial_i g^{(1)}_{jl}+\partial_j g^{(1)}_{il}-\partial_l g^{(1)}_{ij}\right)\Big] .
\end{align}
\end{widetext}
Correspondingly, the Laplace-Beltrami operator can be split as
\begin{align}
\Delta\simeq\Delta^{(0)}+\epsilon \Delta^{(1)}+\epsilon^2 \Delta^{(2)} ,
\end{align}
with the unperturbed operator $\Delta^{(0)}$ (on a flat background this is just the Laplacian in spherical coordinates) and the higher order corrections
\begin{align}
\Delta^{(1)}	=	&-g_{(1)}^{ij}\partial_i\partial_j-\left (g_{(0)}^{ij}\prescript{(1)}{}{\Gamma_{ij}^k}-g_{(1)}^{ij}\prescript{(0)}{}{\Gamma_{ij}^k}\right)\partial_k ,\\
\Delta^{(2)}	=	&\left(g_{(1)}^{ik}g^{(1)j}_{k}-g_{(2)}^{ij}\right)\partial_i\partial_j-\Big[g_{(0)}^{ij}\prescript{(2)}{}{\Gamma_{ij}^k}\nonumber\\
								&+\left(g_{(1)}^{ik}g^{(1)j}_{k}-g_{(2)}^{ij}\right)\prescript{(0)}{}{\Gamma_{ij}^k}-g_{(1)}^{ij}\prescript{(1)}{}{\Gamma_{ij}^k}\Big]\partial_k
.\label{pertlap}
\end{align}
If coordinate sets are employed which naturally describe the underlying metric, the Dirichlet boundary conditions most probably become $\epsilon$-dependent as well (due to the $\epsilon$-dependence of $\partial B_\rho$) which makes a perturbative treatment unnecessarily complicated. This problem can be circumvented employing geodesic coordinates \eqref{geod_coord}. As described above, then $\D \sigma$ denotes the geodesic radial form with respect to the origin $p_0$ and therefore describes the normals of surfaces of constant $\sigma$ while $\D\chi$ and $\D\gamma$ are the corresponding tangents.

On a flat background the \nth{0} order approximation to the metric is just flat space in spherical coordinates as expected and the boundary condition \eqref{boundarycond} has become homogeneous thus simplifying the calculation enormously.

Now the problem can be separated order by order expanding the eigenvalues and -functions as
\begin{align}
\lambda	&=\lambda^{(0)}+\epsilon \lambda^{(1)}+\epsilon^2\lambda^{(2)} ,\\
\psi			&=\psi^{(0)}+\epsilon \psi^{(1)}+\epsilon^2 \psi^{(2)}.
\end{align} 
There is a complication though: Perturbing the metric goes hand in hand with perturbing its determinant which is part of the measure $\D\mu$ with respect to which we define the scalar product of the Hilbert space of our problem $\hil.$ In fact, $\D\mu=\sqrt{g}\D^3x$ but the square root of the determinant of the metric reads
\begin{align}
\sqrt{g}=\sqrt{g^{(0)}}\left[1+\frac{\epsilon}{2}g^{(1)}_{ij}g_{(0)}^{ij}+\Ord(\epsilon^2)\right].
\end{align}
This translates to a perturbative measure $\D\mu=\D\mu^{(0)}+\epsilon\D\mu^{(1)}+\Ord(\epsilon^2)$ which in turn leads to a perturbation of the Hilbert space scalar product $\braket{,}=\braket{,}_0+\epsilon\braket{,}_1+\Ord(\epsilon^2).$

The unperturbed operator $\Delta^{(0)}$ is self-adjoint  with respect to $\D\mu_0$ \ie it is only almost (meaning up to higher order corrections) self-adjoint with respect to $\D\mu.$ Thus, the solutions of the unperturbed eigenvalue problem
\begin{align}
\left(\Delta^{(0)}+\lambda_n^{(0)}\right)\psi_n^{(0)}=0
\end{align}
only furnish an almost orthonormal base of $\hil.$

How to deal with this unusual kind of perturbation is explained in appendix \ref{app_pert}. There this method is applied to an abstract operator $\op$ which in this case corresponds to $\op=-\Delta.$ 

In the main text we will content ourselves with giving the perturbative corrections to the eigenvalues. To \nth{1} order we obtain
\begin{align}
\lambda_n^{(1)}	&=-\braket{\psi_n^{(0)}|\Delta^{(1)}\psi_n^{(0)}}_0\\
							&=-\int\D\mu_0\psi_n^{(0)\dagger}\Delta^{(1)}\psi_n^{(0)}.\label{eig1}
\end{align}
which resembles the usual result. Note that the expectation value is taken with respect to the unperturbed measure. There is a significant change, though, to \nth{2} order. The corresponding correction reads
\begin{widetext}
\begin{align}
\lambda_n^{(2)}=	&-\braket{\psi_n^{(0)}|\Delta^{(2)}\psi_n^{(0)}}_0-\sum_{m\neq n}\frac{\braket{\psi_m^{(0)}|\Delta^{(1)}\psi_n^{(0)}}_0\braket{\psi_n^{(0)}|\Delta^{(1)}\psi_m^{(0)}}_0}{\lambda_m^{(0)}-\lambda_n^{(0)}}\nonumber\\
							&-\braket{\psi_n^{(0)}|\Delta^{(1)}\psi_n^{(0)}}_1+\sum_{m}\braket{\psi_m^{(0)}|\Delta^{(1)}\psi_n^{(0)}}_0\braket{\psi_n^{(0)}|\psi_m^{(0)}}_1.\label{eig2}
\end{align}
\end{widetext}

Then, restoring all quantum numbers $nlm$ the momentum standard deviation of the state $\psi_{nlm}$ becomes to \nth{2} order in $\epsilon$
\begin{align}
\sigma_p	&=\hbar\sqrt{\braket{-\Delta}} \nonumber \\
				&=\hbar\sqrt{\lambda^{(0)}_{nlm}+\epsilon\lambda^{(1)}_{nlm}+\epsilon^2\lambda^{(2)}_{nlm}}.
\end{align}

In principle this approach can be used to get perturbative corrections to any already known solution. Nevertheless, we will concentrate on flat backgrounds.

\subsection{Flat background perturbations}

In this paper we are particularly interested in perturbing around a flat background \ie flat space in spherical coordinates constructed around $p_0.$ Then, the \nth{0} order problem has general solutions characterized by the integers $n,l,m:$
\begin{align}
\psi^{(0)}_{nlm}	&=\sqrt{\frac{2}{\rho^3j^2_{l+1}(j_{l,n})}}j_{l}\left(j_{l,n}\frac{\sigma}{\rho}\right)Y^l_m(\theta,\phi) ,\label{0sol}\\
\lambda^{(0)}_{nlm}		&=\left(\frac{j_{l,n}}{\rho}\right)^2 ,
\end{align}
with the spherical harmonics $Y_m^l,$ the spherical Bessel function of \nth{1} kind $j_l(x)$ and the $\text{n}^{\text{th}}$ zero of the spherical Bessel function of \nth{1} kind $j_{l,n}.$ 

It is well-known that the lowest eigenvalue of the unperturbed operator equals $\lambda_{100}=\pi^2/\rho^2.$ Thus, the uncertainty acquires the lower limit
\begin{align}
\sigma_p	&\gtrsim \frac{\hbar\pi}{\rho}\sqrt{1+\left(\epsilon\lambda^{(1)}_{100}+\epsilon^2\lambda^{(2)}_{100}\right) \frac{\rho^2}{\pi^2}}.
\end{align}

The state in question is then described by the wave function
\begin{align}
\psi^{(0)}_{100}=\frac{1}{\sqrt{2\pi\rho}}\frac{\sin\left(\pi\frac{\sigma}{\rho}\right)}{\sigma} ,
\end{align}
which is independent of $\chi$ and $\gamma.$ Using this and the characteristic of the geodesic coordinates \eqref{geod_met_char}, the relevant terms of the Laplace-Beltrami operator (\ie the ones corresponding to derivatives with respect to $\sigma$ denoted as $\Delta^{(n)}_\sigma$) become order by order
\begin{align}
\Delta_\sigma^{(1)}=	&\frac{1}{2}\left(g_{(0)}^{ij}\partial_\sigma g^{(1)}_{ij}-g_{(1)}^{ij}\partial_\sigma g^{(0)}_{ij}\right)\partial_\sigma ,\\
\Delta_\sigma^{(2)}=	&\frac{1}{2}\Big[g_{(0)}^{ij}\partial_\sigma g^{(2)}_{ij}-g_{(1)}^{ij}\partial_\sigma g^{(1)}_{ij}\\
											&+\left(g_{(1)}^{ik}g^{(1)j}_{k}-g_{(2)}^{ij}\right)\partial_\sigma g^{(0)}_{ij}\Big]\partial_\sigma.
\end{align}
Thus, the problem is much more tractable than it appears at first glance.

This is how far we can get without additional assumptions about the problem. These will be made in the following section.

\section{General solution for small position uncertainties\label{gensol}}

The only assumption that this calculation is based on demands that the geodesic balls be sufficiently small \ie that we stay in a coordinate neighbourhood of $p_0$ which is small enough such that we can approximate any metric in terms of Riemann normal coordinates $x^a$ as \cite{RNC}
\begin{align}
g_{ij}\simeq e^a_ie^b_j\left(\delta_{ab}-\frac{1}{3}R_{acbd}|_{p_0}x^{c}x^{d}\right)
\end{align}
with the \emph{dreibein} $e^a_i$ ($i,j,\dots$ denote spherical coordinates $\sigma_0,$ $\chi_0$ and $\gamma_0$ constructed around the point $p_0$) and the Riemann tensor $R_{acbd}.$ Thus, to \nth{0} approximation space is flat and there is no first order correction.

Correspondingly, the inverse metric reads
\begin{align}
g^{ij}		&\simeq e^i_ae^j_b\left(\delta^{ab}+\frac{1}{3}R^{a~b}_{~c~d}|_{p_0}x^{c}x^{d}\right)
\end{align}
where the flat space metric $\delta_{ab}$ is used to rise and lower indices.

As the \nth{1} order correction to the metric vanishes, the \nth{2} order can be treated similarly to a \nth{1} order correction.

Firstly, it will be shown that Riemann normal coordinates are "geodesic Cartesian coordinates" in the sense that spherical coordinates $\sigma_0^i$ constructed around the center of the balls $p_0$ already equal geodesic coordinates $\sigma^i$ (\cf \eqref{geod_coord}). This simplifies the problem enormously thus opening up the possibility to solve the general problem analytically.

In fact, it suffices to show that \eqref{geod_met_char} is satisfied. In general, we have
\begin{align}
g_{\sigma_0 i}\simeq e^a_{\sigma_0}e^b_j\left(\delta_{ab}-\frac{1}{3}R_{acbd}|_{p_0}x^{c}x^{d}\right).
\end{align}
Note that we can assign $x^a=\sigma_0 l^a$ with the unit radial vector
\begin{align}
l^a=(\sin\chi_0 \cos\gamma_0,\sin\chi_0 \sin\gamma_0,\cos\chi_0)
\end{align}
and similarly, $e_{\sigma_0}^a=l^a.$

Plugging this in, we obtain
\begin{align}
g_{\sigma_0 i}	&\simeq l^a e^b_i\left(\delta_{ab}-\frac{\sigma_0^2}{3}R_{acbd}|_{p_0}l^{c}l^{d}\right)
\end{align}
which by the symmetries of the Riemann tensor and the orthonormality of spherical coordinates becomes
\begin{align}
g_{\sigma_0 i}	&=l_a e^a_i\\
						&=\delta_i^{\sigma_0}.
\end{align}
In particular, the same applies for higher order corrections showing that Riemann normal coordinates are in fact geodesic Cartesian coordinates. This makes an additional change of coordinates unnecessary and we can simply to continue with the formalism developed above.

The radial part of the \nth{2} order Laplace-Beltrami-operator reads in this case
\begin{align}
\Delta^{(2)}_\sigma		&=-\frac{\sigma}{3}R_{acbd}|_{p_0}\delta^{ab}l^cl^d\partial_\sigma\\
									&=-\frac{\sigma}{3}R_{cd}|_{p_0}l^cl^d\partial_\sigma
\end{align}
where the \nth{2} equality is valid because the metric at $p_0$ is in fact the identity.
This translates into the \nth{2} order correction to the eigenvalue
\begin{align}
\lambda_{100}^{(2)}	&=-\braket{\psi_{100}^{(0)}|\Delta^{(2)}\psi_{(100)}}_0\nonumber\\
								&=\frac{1}{3}\int_0^\rho \sigma^3\psi_{100}^{(0)\dagger}\partial_\sigma\psi_{100}^{(0)}\int_{S^2}R_{ab}|_{p_0}l^al^b\D\Omega
\end{align}
with the area element of the geodesic two-sphere $\D\Omega.$ The two integrals can be evaluated independently to yield
\begin{align}
\int_0^\rho \sigma^3\psi_{100}^{(0)\dagger}\partial_\sigma\psi_{100}^{(0)}	&=-\frac{3}{8\pi}\\
\int_{S^2}R_{ab}|_{p_0}l^al^b\D\Omega	&=\frac{4\pi}{3}R_{ab}|_{p_0}\delta^{ab}\nonumber\\
																&=\frac{4\pi}{3}R|_{p_0}.
\end{align}
Hence, we obtain the \nth{2} order correction
\begin{align}
\lambda_{100}^{(2)}=-\frac{R|_{p_0}}{6}
\end{align}
which leads to the uncertainty relation
\begin{align}
\sigma_p\rho	&\gtrsim \pi\hbar\sqrt{1-\frac{R|_{p_0}}{6\pi^2} \rho^2}\\
						&\simeq \pi\hbar\left(1-\frac{R|_{p_0}}{12\pi^2} \rho^2 \right).\label{res}
\end{align}

A higher order treatment of the problem is possible, yet tedious. An account to \nth{4} order is given in appendix \ref{HOCORR}. In short, the \nth{3} order correction vanishes and plugging in \eqref{res4app} the final formula for the uncertainty up to 4th order reads
\begin{align}
\sigma_p\rho	\gtrsim	& \pi\hbar\sqrt{1-\frac{R\mid_{p_0}}{6\pi^2}\rho^2- \frac{\xi}{\pi^2}\left(\Psi_2^{~2}+\frac{\Delta R}{15}\right)\Bigg|_{p_0} \rho^4}
\label{res4}
\end{align}
with the \nth{0} order Cartan invariant $\Psi_2$ which satisfies in three dimensions \cite{CartanInvariant}
\begin{align}
\Psi_2^{~2}=\frac{3R^{ab}R_{ab}-R^2}{72} ,
\end{align}
and the numerical constant
\begin{align}
\xi=\frac{2\pi^2-3}{8\pi^2}.
\end{align} 

Hence, the uncertainty relation is also altered on manifolds with vanishing Ricci scalar such as hypersurfaces of constant time in the static Schwarzschild patch (\cf Section \ref{sphapp}). Yet, the influence is significantly weakened.  

\section{Non-uniform applications} 
\label{applic}

In the preceding sections we introduced a formalism that makes an analytical treatment of the EUP possible without assuming further symmetries.  Now the result will be applied to several examples. We will examine spherically symmetric metrics (constant curvature, spatial sections of Schwarzschild, Schwarzschild-(anti-)de Sitter, ''hairy'' and extremal Reissner-Nordstr\"om black holes) and spacelike hypersurfaces of the G\"odel universe.

\subsection{Spherical symmetry\label{sphapp}}

Written in terms of the coordinates $r^i=(r,\theta,\phi)$ in Lema\^itre-Tolman-Bondi form \cite{LTB}, every spherically symmetric line element can be expressed as
\begin{align}
\D s^2=\frac{\D r^2}{1+2E(r)}+r^2\left(\D\theta^2+\sin^2\theta\D\phi^2\right).
\end{align}
The form of the function $E(r)$ indicates that we deal with {\it spatially dependent} curvature. This ansatz was used in many cosmological attempts to model dark energy due to a spherically symmetric void of matter in the universe \cite{void} leading to a dipolar distribution of objects in the sky \cite{dipole}. 

We denote the distance of the geodesic ball from the center of symmetry by $r_0.$ Due to the symmetry of the background space-time this completely characterizes the position of the center of the ball $p_0$ because the curvature function and its derivatives satisfy $E^{(n)}_0\equiv E^{(n)}|_{p_0}=E^{(n)}(r_0)$ where the subscript in this case denotes the $\text{n}^{\text{th}}$ derivative with respect to $r.$

First, say $R|_{p_0}\neq 0.$ Then, the leading order contribution reads
\begin{align}
\sigma_p\rho	&\geq \pi\sqrt{1+2\frac{E_0+r_0 E'_0}{3\pi^2r_0^2} \rho^2}\\
						&\simeq \pi\left(1+\frac{E_0+r_0E'_0}{3\pi^2r_0^2} \rho^2 \right).\label{sphleadcont1}
\end{align}
Plugging in \eg $E(r)=(-1/2)Kr^2,$ that is assuming constant curvature $K$, this equation recovers the result that was already obtained non-perturbatively in Ref. \cite{Schuermann2018} and stated in \eqref{Kgeom}. Additionally, the \nth{4} order contribution vanishes thereby validating the formalism - a perturbative expansion of a polynomial trivially equals itself confirming the result cited above.

Secondly and conversely, the relation $R|_{p_0}=0$ is only satisfied by constant-time sections of the static Schwarzschild patch. The corresponding result describes the uncertainty relation in the presence of massive bodies (\eg on the surface of planets or outside the horizon of black holes). It becomes to \nth{4} order
\begin{align}
\sigma_p\rho\gtrsim 	&\pi\hbar\sqrt{1-\frac{\xi M^2 }{4\pi^2r_0^6} \rho^4}\label{sphleadcont2}\\
\simeq							&\pi\hbar\left(1-\frac{\xi M^2}{8\pi^2r_0^6} \rho^4\right).
\end{align}

Combining the results, the uncertainty relation for the spatial sections of the static patch of the Schwarzschild-(anti-)de Sitter-space-time can be obtained. The corresponding curvature function reads
\begin{align}
E(r)=-\frac{Kr^2}{2}-\frac{M}{r}
\end{align}
where $K=\Lambda/3$ with the cosmological constant $\Lambda$ which can be expressed in terms of curvature scales as $\Lambda/3=R_H^2$ in dS (where it is interpreted as the Hubble radius) and $\Lambda/3=-R_H^2$ in AdS. Correspondingly, the uncertainty relation can be expressed as
\begin{align}
\sigma_p\rho\gtrsim 	&\pi\hbar\sqrt{1-\frac{K}{\pi^2} \rho^2 -\frac{\xi M^2 }{4\pi^2r_0^6} \rho^4}\\
\simeq							&\pi\hbar\left(1-\frac{K}{2\pi^2}\rho^2-\frac{\xi M^2 }{8\pi^2r_0^6}\rho^4\right).
\end{align}
This equation also indicates the expected position at which both effects are equally strong. This happens when
\begin{align}
r_0=\sqrt[6]{\frac{\xi M^2\rho^2}{4K}}\sim\sqrt[3]{ R_S R_H \rho}
\end{align} 
with $R_S =2M.$ If $R_H\gg R_S,$ in the de Sitter-case these quantities still describe the positions of the horizons accurately. Thus, the relative strength of the effects depends on the position uncertainty.

Practically, the constant-time slices of our universe are not or at least just very slightly curved. Say the radius of curvature is of the order of the Hubble radius of our universe and an object is situated in the vicinity of the earth for which $R_S\sim 10\text{mm}.$ Additionally assume (as for Hawking radiation) that $\rho\sim R_S.$ Then the earth's influence is stronger than the constant curvature influence up to a distance of $r_0\sim 10^4\text{km}$ from the earth's center \ie in the atmosphere. Hence, for laboratory experiments on the surface of the earth both effects need to be taken into account.

For Sagittarius A*, the black hole in the center of the Milky Way, this distance has already grown to $r_0\sim 10^4\text{AU}\sim 1\text{ly}$ which extends far out to the orbits of the closest stars. 

This could lead to interesting results in neutron-star and near black-hole physics. The Chandrasekhar mass would be one of such applications (see \cite{Neutron}). 

Another possible extension are the spatial sections of a simple model of a hairy black hole based on the energy function $E(r)=-M/r+Q^{2\omega}/r^{2\omega}$ \cite{hairy} with the constant ''hairy charge'' $Q$ and the particle number per unit area ansatz parameter $\omega$. This geometry is a black hole solution of Einstein's equations in 4-dimensions with an anisotropic fluid stress energy tensor. For $\omega=1$ it describes the Reissner-Norstr\"om metric, while it develops short hair for $\omega>1$.

From \eqref{sphleadcont1} and \eqref{sphleadcont2} we obtain 
\begin{widetext}
\begin{eqnarray}
\sigma_p\rho\geq \pi\hbar\sqrt{1-(1-2\omega)\frac{2}{3\pi^2} \frac{Q^{2\omega}}{r_0^{2(1+\omega)}}  \rho^2 - 
\frac{1}{4\pi^2} \frac{\xi M^{2}}{r_0^6} \rho^4 + \Ord_{\omega}(\rho^4)}
\label{hairy2}
\end{eqnarray}
\end{widetext}
where the $\Ord(\rho^4)$ contribution for the hairy charge part was omitted for simplicity. This can be done for $Q\ll M$ and $\omega\geq1.$ 

On the other hand, the result for the extremal Reissner-Nordstr\"om black hole ($Q=M$ and $\omega=1$) can be given in terms of the $\nth{2}$ order contribution as
\begin{align}
\sigma_p\rho\geq \pi\hbar\sqrt{1+\frac{2}{3\pi^2} \frac{M^{2}}{r_0^{4}}\rho^2}.
\end{align}
Thus, in contrast to the Schwarzschild black hole the contribution is positive, thereby increasing the uncertainty. This aligns well with the fact that adding charge adds negative curvature which in turn reflects the relative sign between the charge and mass terms in the energy function.

\subsection{G\"odel universe}

Describing the G\"odel universe on spacelike hypersurfaces of the slicing given in \cite{Goedel,Godel}, the 3-metric reads
\begin{align}
\D s^2=\frac{\D r^2}{1+\left(\frac{r}{2a}\right)^2}+r^2\left[1-\left(\frac{r}{2a}\right)^2\right]\D\phi^2+\D z^2
\end{align}
with the rotation radius $a$ which is related to the angular velocity as $\Omega_G=1/(\sqrt{2}a).$ 

Due to the symmetry of the problem the position of the center of the ball is characterized by the polar distance from the rotation axis $r_0.$

The corresponding leading order uncertainty relation reads
\begin{align}
\sigma_p\rho\geq \pi  \hbar  \sqrt{1-\frac{\rho ^2}{3 \pi ^2 a^2}\left(1-\frac{1}{C}-\frac{1}{2 C^2}\right)}
\end{align}
with
\begin{align}
C=1-\left(\frac{r_0}{2a}\right)^2.
\end{align}

This example concludes this section on non-uniform applications of the asymptotic EUP. However, we have not yet shown how to reconcile this relation with the well-known GUP from the quantum gravity side. This will be done in the following section.

\section{Asymptotic Generalized Extended Uncertainty Principle - AGEUP}

In this section we will shortly comment on how to incorporate quantum gravity based effects into the problem. The (linear and quadratic) generalised uncertainty principle is often implemented in 3-D quantum mechanics  by virtue of a deformed commutator (\cf \cite{Das2008,Ali2009,GUPReview})
\begin{align}
[\hat{x}^i,\hat{P}_j]=	&i\hbar\Big[\delta^i_j+\alpha'\left(\hat{P}\delta^i_j+\frac{g^{ik}\hat{P}_k\hat{P}_j}{P}\right)\nonumber\\
				&+(\tilde{\alpha}-\alpha'^2)\hat{P}^2\delta^i_j+(2\tilde{\alpha}-\alpha'^2)g^{ik}\hat{P}_k\hat{P}_j\Big]
\end{align}
where the parameters $\tilde{\alpha}$ and $\alpha'$ are dimensionful. More general deformed commutators which involve the coordinates are also possible \cite{Mignemi2019}, but we will not consider them here.

We will only take into account the quadratic GUP here thus setting $\alpha'=0$ and denote suggestively $\tilde{\alpha}=\alpha l_p^2/\hbar^2$ with the dimensionless GUP-parameter $\alpha$ and the Planck length $l_p.$ 

Thus, the deformed commutator becomes
\begin{align}
[\hat{x}^i,\hat{P}_j]=i\hbar\left[\delta^i_j\left(1+\frac{\alpha l_p^2}{\hbar^2}\hat{P}^2\right)+2\frac{\alpha l_p^2}{\hbar^2}g^{ik}\hat{P}_k\hat{P}_j\right].\label{defcom}
\end{align}
With this choice of parameters noncommutativity of space coordinates appears only at order $l_p^4\hat{p}^4/\hbar^4$ and can thus be neglected.

We can write the momentum operator in terms of an auxiliary quantity $\hat{p}_i$ which satisfies the usual commutation relations $[\hat{x}^i,\hat{p_j}]=i\hbar$ \ie it represents the normal quantum mechanical momentum operator. Following this ansatz, the thus called deformed momentum operator $\hat{P}_i$ reads
\begin{align}
\hat{P}_i=\hat{p}_i\left(1+\frac{\alpha l_p^2}{\hbar^2}\hat{p}^2\right).
\end{align}

All the results obtained in this paper were written in terms of eigenvalues of $\hat{p}^2.$ As $\hat{p}_i$ and $\hat{P}_i$ commute, we can simply translate the results using $\hat{P}_i$ terms of the results with respect to $\hat{p}_i.$ This works as follows:

\emph{Per definitionem} we have
\begin{align}
\sigma_P=\sqrt{\braket{\hat{P^2}}-g^{ij}\braket{\hat{P}_i}\braket{\hat{P}_j}}.
\end{align}
As the wave-functions solving the eigenvalue problem \eqref{evp} furnish an orthonormal basis of the Hilbert space in question, we can write
\begin{align}
\braket{P_i}	&=\braket{\psi|\hat{P}_i|\psi}\\
					&=\braket{\psi|\hat{p}_i\left(1+\frac{\alpha l_p^2}{\hbar^2}\hat{p}^2\right)|\psi}\\
					&=\braket{\psi|\hat{p}_i\left(1+\alpha l_p^2\lambda^2\right)|\psi}\\
					&=\braket{\hat{p}_i}\left(1+\alpha l_p^2\lambda^2\right)\\
					&=0
\end{align}
where we used \eqref{evp} and the fact that $\braket{\hat{p}_i}=0$ within the ball \cite{Schuermann2018}.

On the other hand, we have
\begin{align}
\braket{\hat{P^2}}	&=\braket{\psi|\hat{p^2}\left(1+2\frac{\alpha l_p^2}{\hbar^2}\hat{p}^2\right)|\psi}+\mathcal{O}\left(\frac{l_p^4|\hat{p}|^4}{\hbar^4}\right)\\
							&\simeq\hbar^2\lambda\left(1+2\alpha l_p^2\lambda\right).
\end{align}
Hence, we obtain the uncertainty relation
\begin{align}
\sigma_P\rho	&\geq\hbar\rho\sqrt{\lambda\left(1+2\alpha l_p^2\lambda\right)}\\
						&\gtrsim\hbar\sqrt{\rho^2\lambda}\left(1+\alpha l_p^2\lambda\right).
\end{align}
Plugging in the eigenvalue in the general case to lowest non-vanishing order of \eqref{res4app} leads to the asymptotic form of both the EUP and the GUP, \ie to the AGEUP
\begin{align}
\sigma_P\rho\gtrsim	&\pi\hbar\sqrt{1-\frac{R|_{p_0}\rho^2}{6\pi^2}+\Ord (\rho^4)}\nonumber\\
									&\times\left[1+\pi^2\alpha\frac{l_p^2}{\rho^2}\left(1-\frac{R|_{p_0}\rho^2}{6\pi^2}+\Ord (\rho^4)\right)\right]\\
				\simeq			&\pi\hbar\sqrt{1-\frac{R|_{p_0}\rho^2}{6\pi^2}+\Ord (\rho^4)}\left(1+\alpha\frac{l_p^2\sigma_p^2}{\hbar^2}\right)\\
				\simeq 		&\pi\hbar\left(1-\frac{R|_{p_0}\rho^2}{12\pi^2}+\alpha\frac{l_p^2\sigma_p^2}{\hbar^2}+\Ord (\rho^4)\right)\label{derived_GEUP}
\end{align}
for negligible $R|_{p_0}\rho^2$ and $l_p^2/\rho^2.$ In the special case of flat space this recovers the usual GUP:
\begin{align}
\sigma_P\rho	\gtrsim\pi\hbar\left(1+\alpha\frac{l_p^2\sigma_p^2}{\hbar^2}\right).
\end{align}

In every other case it leads to a GEUP in the given space once the EUP is known.

\section{Summary}
\label{Summary} 

We have presented a formalism which allows to derive the Extended Uncertainty Principle (EUP) asymptotically on 3-dimensional Riemannian manifolds of arbitrary curvature as given by our formula \eqref{res4}. The result represents the uncertainty an observer (defining the foliation of spacelike 3-hypersurfaces) measures on a curved background. The formalism has first been applied to already derived homogeneous and isotropic space models, to be subsequently broadened in scope by inclusion of models which allow for other curvature related effects such as inhomogeneities, rotation and an anisotropic fluid leading to black hole ''hair''. The corresponding results influence neutron star and near black hole physics as well as cosmology, opening up an avenue for further investigation by observations.

Interestingly, the leading (\nth{2} order) curvature induced correction is proportional to the Ricci scalar, while the \nth{4} order correction is proportional to the \nth{0} order Cartan invariant $\Psi_2$ and the curved space Laplacian of the Ricci scalar all evaluated at the expectation value of the position operator.  We have shown that in the case of vanishing Ricci curvature (\eg for the Schwarzschild metric), the lowest order perturbation is of \nth{4} order, while for the rotating G\"odel metric as well as for anisotropic stress ''hairy'' black holes it is already of \nth{2} order, though higher-order terms contribute as well. 

Additionally, we have compared the influence which is exerted upon the uncertainty relation by cosmological horizons to the ones caused by local massive objects (planets, stars, black holes and so on) and proved the necessity to take into account the latter in many sensible applications on small (Earth) and large (Sagittarius A*) scales.

Finally, the formalism has been extended phenomenologically to combine the result with the Generalized Uncertainty Principle (GUP) by virtue of deformed commutators. Thus, we have presented an asymptotic form of the Generalized Extended Uncertainty Principle in the low energy limit given by the formula \eqref{derived_GEUP} which is our main achievement.

\appendix
\section{Measure changing non-singular perturbation theory to \nth{2} order\label{app_pert}}

Consider an operator $\op$ with domain $D$ acting on the Hilbert space $\hil=L^2(D\subseteq{\rm I\!R^3,}\D\mu)$ which is self-adjoint with respect to the measure $\D\mu.$ Then it satisfies the eigenvalue equation
\begin{align}
\op \psi_n	&=\lambda_n\psi_n~\text{in~} D\label{ev}\\
\psi_n			&=0~~~\text{on~}\partial D
\end{align}
with its eigenvalues $\lambda_n$ and eigenvectors $\psi_n$ ($n$ may stand for multiple quantum numbers) which are assumed to be normalized such that
\begin{align}
\braket{\psi_n|\psi_n}\equiv\int_D\D\mu\psi_n^\dagger\psi_n=1.\label{norm}
\end{align}

Further assume that the operator can be expanded perturbatively as
\begin{align}
\op			&=\sum_{N=0}^\infty\epsilon^N\op^{(N)}
\end{align}
with $\epsilon\ll1$ and where all operators $\op^{(N)}$ are $\op^{(0)}$-bounded. The latter is then called the unperturbed operator whose eigenvalues and -vectors are known. This is the starting point for perturbation theory in quantum mechanics.

A perturbative treatment of the eigenvalue problem \eqref{ev} can be complicated by the fact that the unperturbed operator might not be self-adjoint with respect to the measure $\mu.$ In fact, it may instead be self-adjoint with respect to a measure $\D\mu_{0}.$ In usual introductions to perturbation theory $\D\mu=\D\mu_0$ is assumed.

Yet, this is not generally the case. In principle, the measures may be related as
\begin{align}
\D\mu(x)=\D\mu_0(x) a^2(x)\label{measure}
\end{align}
with a positive non-vanishing function of the coordinates $a^2(x).$

A simplifying assumption that has to be made here requires that the function $a^2(x)$ may be expanded perturbatively as
\begin{align}
a^2(x)=1+\sum_{N=1}^{\infty}\epsilon^N a_{N}^{2}(x).\label{measass}
\end{align}
It is questionable whether this problem is solvable or even mathematically well-defined without this additional assumption. Anyway, it certainly applies to the physical example studied in section \ref{EP}.

Expanding $a^2(x)$ as in \eqref{measass} is equivalent to introducing a perturbed scalar product as
\begin{align}
\braket{\psi|\phi}=\sum_{N=0}^{\infty}\epsilon^N\braket{\psi|\phi}_N.\label{prodexp}
\end{align}

Additionally, we assume the spectrum of $\op^{(0)}$ to be discrete \ie we consider non-singular perturbation theory. Of course, this could be generalised to the singular case. 

By virtue of \eqref{measass} the unperturbed operator is almost self-adjoint meaning that the non-self-adjointness is of higher order \ie
\begin{align}
\braket{\psi|\op^{(0)}\phi}_0=\braket{\op^{(0)}\psi|\phi}_0.
\end{align}
In fact, the eigenvectors of the unperturbed operator $\psi_n^{(0)}$ can be normalised (without loss of generality) as $\braket{\psi^{(0)}_n|\psi^{(0)}_n}_0=1.$ Thus, they span an orthonormal basis of the Hilbert space corresponding to the unperturbed problem $\hil_0=L^2(D\subseteq{\rm I\!R^3,}\D\mu_0)$ while the eigenvectors of the full operator $\psi_n$ span an orthonormal basis of $\hil.$ Hence, we can expand
\begin{align}
\psi^{(i)}_n	&=\sum_{m}c^{(i)}_{nm}\psi_m\\
					&=\sum_{m}c^{(1)}_{nm}\psi_m^{(0)}+\Ord(\epsilon)
\end{align}
with the  complex $\text{i}^{\text{th}}$ order coefficients $c^{(i)}_{nm}.$
In particular, we approximate
\begin{align}
\psi^{(1)}_n	&\simeq \sum_{m}c^{(1)}_{nm}\psi^{(0)}_m.\label{1stperteigvec}
\end{align}

Furthermore, from \eqref{norm} and the normalization of $\psi^{(0)}_n$, we deduce
\begin{align}
\braket{\psi^{(1)}_n|\psi^{(0)}_n}_0+\braket{\psi^{(0)}_n|\psi^{(1)}_n}_0+\braket{\psi^{(0)}_n|\psi^{(0)}_n}_1	&=0\label{norm1st}
\end{align}
and
\begin{widetext}
\begin{align}
\braket{\psi^{(2)}_n|\psi^{(0)}_n}_0+\braket{\psi^{(0)}_n|\psi^{(2)}_n}_0+\braket{\psi^{(0)}_n|\psi^{(0)}_n}_2+\braket{\psi^{(1)}_n|\psi^{(1)}_n}_0+\braket{\psi^{(1)}_n|\psi^{(0)}_n}_1+\braket{\psi^{(0)}_n|\psi^{(1)}_n}_1&=0\label{norm2nd}.
\end{align}
\end{widetext}

The first of these relations \eqref{norm1st} translates to
\begin{align}
\text{Re}\left(c_{nn}^{(1)}\right)=-\frac{1}{2}\braket{\psi_n^{(0)}|\psi_n^{(0)}}_1
\end{align}
which yields without loss of generality (discarding a global phase)
\begin{align}
c_{nn}^{(1)}=-\frac{1}{2}\braket{\psi_n^{(0)}|\psi_n^{(0)}}_1.\label{coeffnn}
\end{align}
At this point we see the first influence of the non-self-adjointness of $\op^{(0)}:$ The first order eigenvectors cannot be taken to be orthogonal to the unperturbed ones because they themselves are only orthogonal to \nth{0} order \ie
\begin{align}
\braket{\psi_n^{(0)}|\psi_m^{(0)}}=\delta_{nm}+\Ord(\epsilon).\label{almorth}
\end{align}
Perturbing \eqref{ev} yields order by order
\begin{widetext}
\begin{align}
\mathcal{O}(1)&
\begin{cases}
\left(\op^{(0)}-\lambda_n^{(0)}\right)	&\psi_n^{(0)}=0|_{D}\\
																		&\psi_n^{(0)}=0|_{\partial D}
\end{cases}\label{eig0th}\\
\mathcal{O}(\epsilon)&
\begin{cases}
\left(\op^{(0)}-\lambda^{(0)}_n\right)	&\psi_n^{(1)}=-\left(\op^{(1)}-\lambda_n^{(1)}\right)\psi_n^{(0)}\Big|_{D}\\
																		&\psi_n^{(1)}=0|_{\partial D}
\end{cases}\label{eig1st} \\
\mathcal{O}(\epsilon^2)&
\begin{cases}
\left(\op^{(0)}-\lambda^{(0)}_n\right)	&\psi_n^{(2)}=-\left(\op^{(2)}-\lambda^{(2)}_n\right)\psi_n^{(0)}-\left(\op^{(1)}-\lambda_n^{(1)}\right)\psi_n^{(1)}\Big|_{D} \\
																		&\psi_n^{(2)}=0|_{\partial D}.
\end{cases}  \label{eig2nd}
\end{align}
\end{widetext}

These equations will now be contracted with $\bra{\psi_m}$ for some $m.$ The case $m=n$ is yields the corrections to the eigenvalues while the coefficients $c^{(1)}_{nm}$ can be determined from the opposite case Note that said contraction shifts the order of perturbation of some terms not only due to the expansion of $\psi_m$ but also of the scalar product. 

The \nth{0} order part then really corresponds to solving the unperturbed eigenvalue problem because
\begin{align}
\lambda_n^{(0)}=\braket{\psi^{(0)}_n|\op^{(0)}\psi_n^{(0)}}_0
\end{align}
takes place entirely within $\hil_0.$

Furthermore, the \nth{1} order corrections to the eigenvalues read (applying \eqref{eig0th})
\begin{align}
\lambda_n^{(1)}=	&\braket{\psi_n^{(1)}|\op^{(0)}\psi_n^{(0)}}_0+\braket{\psi_n^{(0)}|\op^{(0)}\psi_n^{(1)}}_0\nonumber\\
							&+\braket{\psi_n^{(0)}|\op^{(0)}\psi_n^{(0)}}_1+\braket{\psi_n^{(0)}|\op^{(1)}\psi_n^{(0)}}_0\\
						=	&\lambda^{(0)}\Big(\braket{\psi_n^{(1)}|\psi_n^{(0)}}_0+\braket{\psi_n^{(0)}|\psi_n^{(1)}}_0\nonumber\\
							&+\braket{\psi_n^{(0)}|\psi_n^{(0)}}_1\Big)+\braket{\psi_n^{(0)}|\op^{(1)}\psi_n^{(0)}}_0
\end{align}
which with \eqref{norm1st} becomes
\begin{align}
\lambda_n^{(1)}=\braket{\psi_n^{(0)}|\op^{(1)}\psi_n^{(0)}}_0.
\end{align}
Hence, $\op^{(0)}$ not being self-adjoint does not significantly change the \nth{1} order contribution to the eigenvalues. However, the expectation value is taken with respect to the measure $\D\mu_0.$

Contracting the eigenvalue problem with $\bra{\psi_m}$ assuming $n\neq m$ yields (after some algebra and applying \eqref{eig1st})
\begin{align}
c_{nm}^{(1)}=\frac{\braket{\psi_m^{(0)}|\op^{(1)}\psi_n^{(0)}}_0}{\lambda_n^{(0)}-\lambda_m^{(0)}}~\text{for}~n\neq m.\label{coeffnm}
\end{align}
Again, this is equivalent to the case where $\op^{(0)}$ is self-adjoint and again the amplitude is calculated with respect to the unperturbed measure. Thus, to \nth{1} order the only clearly visible change is in the coefficient $c_{nn}^{(1)}$ according to \eqref{coeffnn}.

Analogously, the \nth{2} order correction to the eigenvalues reads after application of \eqref{norm2nd} and \eqref{eig0th}
\begin{widetext}
\begin{align}
\lambda_n^{(2)}=	&\braket{\psi_n^{(0)}|\op^{(2)}\psi_n^{(0)}}_0+\braket{\psi_n^{(0)}|\op^{(1)}\psi_n^{(0)}}_1+\braket{\psi_n^{(1)}|\op^{(1)}\psi_n^{(0)}}_0+\braket{\psi_n^{(0)}|\op^{(1)}\psi_n^{(1)}}_0\nonumber\\
							&+\braket{\psi_n^{(1)}|\left(\op^{(0)}-\lambda_n^{(0)}\right)\psi_n^{(1)}}_0+\braket{\psi_n^{(0)}|\left(\op^{(0)}-\lambda_n^{(0)}\right)\psi_n^{(1)}}_1.
\end{align}
\end{widetext}
Finally, expanding $\psi_n^{(1)}$ as in \eqref{1stperteigvec} and applying \eqref{coeffnn}, \eqref{almorth} and \eqref{coeffnm}, this can be expressed in terms of already known quantities as
\begin{widetext}
\begin{align}
\lambda_n^{(2)}=	&\braket{\psi_n^{(0)}|\op^{(2)}\psi_n^{(0)}}_0+\sum_{m\neq n}\frac{\braket{\psi_m^{(0)}|\op^{(1)}\psi_n^{(0)}}_0\braket{\psi_n^{(0)}|\op^{(1)}\psi_m^{(0)}}_0}{\lambda_n^{(0)}-\lambda_m^{(0)}}+\braket{\psi_n^{(0)}|\op^{(1)}\psi_n^{(0)}}_1-\sum_{m}\braket{\psi_m^{(0)}|\op^{(1)}\psi_n^{(0)}}_0\braket{\psi_n^{(0)}|\psi_m^{(0)}}_1.\label{gen_eig2}
\end{align}
\end{widetext}
The first two terms of this equation appear in usual perturbation theory as well (which is indicated by the fact that they are evaluated with respect to the unperturbed measure) while the last two are entirely new a fact that can be inferred from the higher order scalar products they are featuring. Thus, the change in the measure has a strong effect on the \nth{2} order eigenvalues and cannot be neglected.

\section{Higher order corrections to the asymptotic uncertainty relation} 
\label{HOCORR}

In this appendix we expand upon the perturbative derivation of the asymptotic uncertainty relation in the main text to get to \nth{4} order thereby explaining the origin of the mathematical constant $\xi$ introduced in section \ref{gensol}. The required math is in principle the same as in the text although the calculations get rather lengthy and can in their entirety only be performed numerically. The Riemann normal coordinate expansion of the metric reads to \nth{4} order \cite{RNC}
\begin{widetext}
\begin{align}
g_{ij}\simeq e^a_ie^b_j\left[\delta_{ab}-\frac{1}{3}R_{acbd}|_{p_0}x^{c}x^{d}-\frac{1}{6}\nabla_e R_{acbd}|_{p_0}x^{c}x^{d}x^e+\left(\frac{2}{45}R_{acdg}R^{~~~g}_{bef}|_{p_0}-\frac{1}{20}\nabla_e\nabla_f R_{acbd}|_{p_0}\right)x^{c}x^{d}x^ex^f\right].\label{geod_met4}
\end{align}
\end{widetext}
Furthermore, the root of the determinant of the metric equals to \nth{2} order
\begin{align}
\sqrt{g}\simeq 1-\frac{1}{6}R_{ab}|_{p_0}x^ax^b
\end{align}
which induces a perturbation in the measure
\begin{align}
\D\mu\simeq\D\mu_0\left(1-\frac{1}{6}R_{ab}|_{p_0}x^ax^b\right)
\end{align}
with $\D\mu_0=\sigma^2\D\sigma\D\Omega.$

As in section \ref{gensol}, the symmetries of the Riemann tensor imply that spherical coordinates constructed around $p_0$ are automatically geodesic coordinates. In other words \eqref{geod_met4} already satisfies \eqref{geod_met_char}.

Accordingly, the higher order contributions to the radial part of the Laplace-Beltrami-operator read
\begin{align}
\Delta_\sigma^{(3)}	&=-\frac{\sigma^2}{4}\nabla_c R_{ab}|_{p_0}l^al^bl^c\partial_\sigma\\
\Delta_\sigma^{(4)}	&=-\frac{\sigma^3}{5}\mathcal{R}_{abcd}|_{p_0}l^al^bl^cl^d\partial_\sigma
\end{align}
where we defined for simplicity
\begin{align}
\mathcal{R}_{abcd}\equiv\frac{1}{9}R^{e~f}_{~a~b}R_{ecfd}+\frac{1}{2}\nabla_c\nabla_dR_{ab}.
\end{align}

As there is no \nth{1} order contribution to the metric. the \nth{3} order contribution can be treated as if it was of \nth{1} order while the \nth{4} order contribution can be treated as if it was of \nth{2} order in the sense of appendix \ref{app_pert}.

Hence, the \nth{3} order contribution reads
\begin{align}
\lambda^{(3)}	&=-\braket{\psi_{100}^{(0)}|\Delta^{(3)}\psi_{100}^{(0)}}_0\\
						&=\int_0^\rho\D\sigma\frac{\sigma^4}{4}\psi_{100}^{(0)}\partial_\sigma\psi_{100}^{(0)} \int_{S^2}\D\Omega \nabla_{c}R_{ab}|_{p_0} l^al^bl^c\\
						&=0
\end{align}
where the last equality stems from the fact that the angular integral vanishes. Thus, there is no \nth{3} order contribution.

However, the \nth{4} order eigenvalue is not as trivial. According to \eqref{gen_eig2}, it reads
\begin{widetext}
\begin{align}
\lambda_{100}^{(4)}=&-\braket{\psi_{100}^{(0)}|\Delta^{(4)}\psi_{100}^{(0)}}_0-\sum_{nlm\neq 100}\frac{\braket{\psi_{nlm}^{(0)}|\Delta^{(2)}\psi_{100}^{(0)}}_0\braket{\psi_{100}^{(0)}|\Delta^{(2)}\psi_{nlm}^{(0)}}_0}{\lambda_{nlm}^{(0)}-\lambda_{100}^{(0)}}\nonumber\\
							&-\braket{\psi_{100}^{(0)}|\Delta^{(2)}\psi_{100}^{(0)}}_2+\sum_{nlm}\braket{\psi_{nlm}^{(0)}|\Delta^{(2)}\psi_{100}^{(0)}}_0\braket{\psi_{100}^{(0)}|\psi_{nlm}^{(0)}}_2.\label{ev4}
\end{align} 
\end{widetext}
In order to be able to compute the values of the terms which are not summed over, we need radial integral
\begin{align}
\int_0^\rho\D\sigma\sigma^5\psi_{100}^{(0)}\partial_\sigma\psi_{100}^{(0)} 	&=\frac{5(3-2\pi^2)}{48\pi^3}
\end{align}
and the angular expression
\begin{align}
\int_{S^2}\D\Omega l^al^bl^cl^d																	&=\frac{4\pi}{15}\left(\delta^{ab}\delta^{cd}+\delta^{ac}\delta^{bd}+\delta^{ad}\delta^{bc}\right).
\end{align}
Furthermore, in three dimensions the Riemann tensor can be expressed in terms of the Ricci tensor and the metric (note that at $p_0$ we have $g_{ab}=\delta_{ab}$) as
\begin{align}
R_{abcd}|_{p_0}=	&\delta_{ac}R_{bd}|_{p_0}-\delta_{ad}R_{bc}|_{p_0}-\delta_{bc}R_{ad}|_{p_0}+\delta_{bd}R_{ac}|_{p_0}\nonumber\\
							&-\frac{R|_{p_0}}{2}\left(\delta_{ac}\delta_{bd}-\delta_{ad}\delta_{bc}\right)
\end{align}
and by the contracted Bianchi identity
\begin{align}
\nabla_c\nabla_dR_{ab}|_{p_0}\left(\delta^{ab}\delta^{cd}+\delta^{ac}\delta^{bd}+\delta^{ad}\delta^{bc}\right)=3\Delta R|_{p_0}.
\end{align}

Combining these, we obtain
\begin{align}
\braket{\psi_{100}^{(0)}|\Delta^{(4)}\psi_{100}^{(0)}}_0=	&\rho^2\frac{3-2\pi^2}{120\pi^2}\Bigg(-\frac{R^2|_{p_0}}{9}\nonumber\\
																					&+\frac{14}{27}R^{ab}R_{ab}|_{p_0}+\Delta R|_{p_0}\Bigg),\\
\braket{\psi_{100}^{(0)}|\Delta^{(2)}\psi_{100}^{(0)}}_2=	&\rho^2\frac{3-2\pi^2}{648\pi^2}\left(R^2|_{p_0}+2R^{ab}R_{ab}|_{p_0}\right).																			
\end{align}

Concerning the sums in \eqref{ev4} both $\braket{\psi_{nlm}^{(0)}|\Delta^{(2)}\psi_{100}^{(0)}}_0$ and $\braket{\psi_{100}^{(0)}|\psi_{nlm}^{(0)}}_2$ only give non-vanishing contributions if $l=m=0$ or $l=2,$ $m=-2,-1,0,1,2.$ 

The exact limits of these sums are not known (the summands are complicated terms of spherical Bessel functions evaluated at special zeroes of other Bessel functions). Yet, they show fast convergence and can thus be obtained numerically. Consider the quantity
\begin{widetext}
\begin{align}
S_{lm}=\sum_{n=2}^{\infty}\left(-\frac{\braket{\psi_{nlm}^{(0)}|\Delta^{(2)}\psi_{100}^{(0)}}_0\braket{\psi_{100}^{(0)}|\Delta^{(2)}\psi_{nlm}^{(0)}}_0}{\lambda_{nlm}^{(0)}-\lambda_{100}^{(0)}}+\braket{\psi_{nlm}^{(0)}|\Delta^{(2)}\psi_{100}^{(0)}}_0\braket{\psi_{100}^{(0)}|\psi_{nlm}^{(0)}}_2\right).
\end{align}
\end{widetext}
Using this notation, we can express the \nth{4} order contribution to the eigenvalue as
\begin{align}
\lambda^{(4)}_{100}=&-\braket{\psi_{100}^{(0)}|\Delta^{(4)}\psi_{100}^{(0)}}_0-\braket{\psi_{100}^{(0)}|\Delta^{(2)}\psi_{100}^{(0)}}_2\nonumber\\
							&+\lambda_{100}^{(2)}\braket{\psi_{100}^{(0)}|\psi_{100}^{(0)}}_2+S_{00}+\sum_{m=-2}^2S_{2m}.\label{ev42}
\end{align}
Computing the remaining terms, we obtain
\begin{align}
\lambda_{100}^{(2)}\braket{\psi_{100}^{(0)}|\psi_{100}^{(0)}}_2	&=\frac{3-2\pi^2}{648\pi^2}R^2|_{p_0}\\
S_{00}																							&=\frac{3-2\pi^2}{1944\pi^2}R^2|_{p_0}\\
S_{20}																							&= -\frac{S_{2c}}{3} (T_{20}^{ab}R_{ab}|_{p_0})^2\\
S_{21}+S_{2-1}																				&= -S_{2c}|T_{21}^{ab}R_{ab}|_{p_0}|^2\\
S_{22}+S_{2-2}																				&= -S_{2c}|T_{20}^{ab}R_{ab}|_{p_0}|^2
\end{align}
with the matrices (which are not tensors!)
\begin{align}
T_{20}^{ab}	&=\text{Diag}(1,1,-2)\label{T20}\\
T_{21}^{ab}	&=
\begin{bmatrix}
0	&	0	&	1\\
0	&	0	&	i\\
1	&	i	&	0\\
\end{bmatrix}\label{T21}\\
T_{22}^{ab}	&=
\begin{bmatrix}
1	&	i	&	0\\
i	&	-1	&	0\\
0	&	0	&	0\\
\end{bmatrix}\label{T22}
\end{align}
and the newly introduced numerical constant $S_{2c}\simeq 6*10^{-3}.$ As the matrices \eqref{T20}, \eqref{T21} and \eqref{T22} are not tensors, the contributions $S_{20},$ $S_{21}$ and $S_{22}$ are not scalars. The reason for this unusual behaviour lies in the fact that $\Delta^{(2)}$ and $\braket{,}_2$ are by themselves not scalar quantities. However, summed up they yield
\begin{align}
\sum_{m=-2}^2S_{2m}=S_{2c}\left(\frac{R^2}{6}-\frac{R^{ab}R_{ab}}{2}\right)
\end{align}
which is a scalar as expected.

Finally, plugging all terms into \eqref{ev42}, we obtain the \nth{4} order contribution to the eigenvalue
\begin{align}
\lambda_{100}^{(4)}=-\rho^2\left[\xi_1\left(3R_{ab}R^{ab}-R^2\right)+\xi_2\Delta R\right]|_{p_0}
\end{align}
where we introduced the numerical constants 
\begin{align}
\xi_1	&\simeq 3.0*10^{-3},\\
\xi_2	&=\frac{2\pi^2-3}{120\pi^2}.
\end{align}
Hence, we can approximately say that
\begin{align}
\xi\equiv 72\xi_1\simeq 15\xi_2.
\end{align}
Furthermore, the result can be expressed in terms of the \nth{0} order Cartan invariant $\Psi_2$ which in three dimensions reads \cite{CartanInvariant}
\begin{align}
\Psi_2^{~2}=\frac{3R^{ab}R_{ab}-R^2}{72}
\end{align}
thus yielding
\begin{align}
\lambda_{100}^{(4)}\simeq -\rho^2\xi\left(\Psi_2^{~2}+\frac{\Delta R}{15}\right)\Bigg|_{p_0}
\label{res4app}
\end{align}
with the numerical constant
\begin{align}
\xi=\frac{2\pi^2-3}{8\pi^2}.
\end{align} 
This concludes our treatment of the extended uncertainty relation to \nth{4} order.

\section*{Acknowledgments}

The work of F.W. was supported by the Polish National Research and Development Center (NCBR) project ''UNIWERSYTET 2.0. --  STREFA KARIERY'', POWR.03.05.00-00-Z064/17-00 (2018-2022).

\end{document}